\documentclass{ijuc}
\usepackage{array}
\usepackage{multirow}
\usepackage{amsmath}
\usepackage{amssymb}
\usepackage[pdftex]{graphicx}

\usepackage{booktabs}
\usepackage{braket}

\usepackage{bbm}
\usepackage{microtype}

\usepackage{textcomp}

\usepackage[pdfusetitle,
 bookmarks=true,bookmarksnumbered=true,bookmarksopen=false,
 breaklinks=false,pdfborder={0 0 0},backref=false,colorlinks=true]{hyperref}
\usepackage{doi}

\usepackage{eepic}
\usepackage{xcolor}

\newcommand{\twoVec}[2]{\begin{pmatrix}#1\\#2\end{pmatrix}}

\begin{document}

\title{A Nuclear Magnetic Resonance Implementation of a Classical Deutsch-Jozsa Algorithm}

\author{Alastair A. Abbott\inst{1}
\email{aabb009@aucklanduni.ac.nz}
\and
Matthias Bechmann\inst{2}
\email{matthias.bechmann@york.ac.uk}
\and
Cristian S. Calude\inst{1}
\email{cristian@cs.auckland.ac.nz}
\and
Angelika Sebald\inst{2}
\email{angelika.sebald@york.ac.uk}
}

\institute{
Department of Computer Science, University of Auckland, Private Bag 92019, Auckland, New Zealand
\and
Department of Chemistry, University of York, YO10 5DD, York, UK
}

\def\received{Received 29 September 2011}

\maketitle

\begin{abstract}
Nuclear magnetic resonance (NMR) has been widely used as a demonstrative medium for showcasing the ability for quantum computations to outperform classical ones. 
A large number of such experiments performed have been implementations of the Deutsch-Jozsa algorithm.
It is known, however, that in some cases the Deutsch-Jozsa problem can be solved classically using as many queries to the black-box as in the quantum solution.
In this paper we describe experiments in which we take the contrasting approach of using NMR as a classical computing medium, treating the nuclear spin vectors classically and utilising an alternative embedding of bits into the physical medium.
This allows us to determine the actual Boolean function computed by the black-box for the $n=1,2$ cases, as opposed to only the nature (balanced or constant) as conventional quantum algorithms do.
Discussion of these experiments leads to some clarification of the complications surrounding the comparison of different quantum algorithms, particularly black-box type algorithms.
\end{abstract}


\maketitle

\section{Introduction}


Nuclear magnetic resonance (NMR) experiments are generally conducted using
bulk samples and hence the manipulating radiofrequency (rf) pulses and the
detection signal have to be regarded in the context of an ensemble average
of the underlying nuclear magnetic spin dynamics. Theoretically, this
situation is successfully dealt with by a density matrix approach. Since,
however, the idea of quantum computation is based on the concept of being
able to manipulate the spin dynamics on the basis of pure quantum spin
states, there have been various attempts at implementing quantum computation
algorithms using the experimental conditions and restrictions of NMR by
adopting pseudo-pure spin state based approaches.

The most commonly implemented quantum algorithm, both in NMR and in general, is the one due to Deutsch
and Josza~\cite{Collins1998,Linden1998,Collins2000,Cory2000,Dorai2000,Kim2000,Marx2000,Arvind2001a,Mahesh2001,Kessel2002,Kumar2002,Das2003,Wei2003,Mangold2004,Fitzsimons2007,Fahmy2008,Gopinath2008}.
The various NMR implementations differ by: the underlying spin quantum numbers 
$S$ ($S=1/2$ or $S>1/2$); the initial spin states (thermal equilibrium state
or pseudo pure state); the algorithmic implementation of the problem
(Collins~\cite{Collins1998} or Cleve~\cite{Cleve:1997aa} defining the number of
qubits necessary to operate a given DJ problem size).

In the context of computation, NMR has in the past been exclusively used for
implementing quantum computations. However, it also has
potential as a classical computing medium, where the rich state space can be
fully utilised to perform classical operations~\cite{Rosello-Merino:2010aa}.
The Deutsch-Jozsa problem, long touted as a simple and key example of the
ability of quantum computing to outperform classical computing, has more
recently been shown to be \emph{de-quantisable} in some cases---i.e.,\ efficient classical solutions can be formulated~\cite%
{Abbott:2011aa,Calude:2007aa}.

In this paper, we describe the implementation of the $n=1$ and $n=2$
de-quantised solutions for the DJ problem in a classical NMR computation.
The process of implementing this solution highlights key aspects of quantum
algorithms and computation, and we discuss these in detail. In
particular, we emphasise the separation between three nested `layers' of any
quantum algorithm: the problem formulation, the algorithm formulation, and
the physical implementation. 
In general these levels are independent, but
certain conditions on the relationship between levels must be satisfied. Specifically, a particular algorithm applies only to a specific problem formulation, and for each algorithm a choice of embedding into the physical medium must be made in order to implement it. 
Further, for `oracle' or `black-box' problems such as the Deutsch-Jozsa problem the comparison of different formulations of the problem requires discussion of the ability to embed the black-box from one formulation into the other. All these issues are subtle and require further discussion.

\section{Problem Formulation}

The standard formulation of the Deutsch-Jozsa problem~\cite{Deutsch:1992aa} is as follows. 
Let $f:\{0,1\}^n \to \{0,1\}$, and suppose we are given a black-box computing $f$ with the guarantee that $f$ is either constant (i.e. for all $x_1,x_2 \in \{0,1\}^n$ we have $f(x_1)=f(x_2)$) or balanced (i.e. $f(x) = 0$ for exactly half of all possible inputs $x \in \{0,1\}^n$). 
The problem is to determine, in as few black-box calls as possible, whether $f$ is constant or balanced. 
The obvious classical algorithm must examine one more than half the input bit-strings and thus requires $2^{n-1}+1$ black-box calls, while the quantum solution requires only one. 

There is, however, an important subtle issue: the classical and quantum problems are slightly different. In one case we are given a \emph{classical} black-box $C_f$ computing $f$, and in the other we are given a \emph{unitary} black-box $U_f$, operating in a $2^{n+1}$ dimensional Hilbert space $\mathcal{H}_{2^{n+1}}$, computing $U_f\ket{x}\ket{y}=\ket{x}\ket{y \oplus f(x)}$ where `$\oplus$' denotes addition modulo 2. As such, it is bending the truth a little to say that `\emph{the} problem' can be solved more efficiently quantum mechanically than classically.

The possibility in the $n=1,2$ cases to de-quantise the quantum solution to give an equally good classical algorithm~\cite{Abbott:2011aa,Calude:2007aa} results from working with a formulation of the problem in which we are given a higher dimensional classical black-box. 
In this case, complex numbers are used as `complex bits'---a classical analogue of a qubit---and the black-box $C_f$ operates (in the $n=2$ case) as follows:
\begin{equation}
	\label{deQuant:eqn:2bitCf}
	C_f\twoVec{z_1}{z_2} = C_f\twoVec{a_1+b_1i}{ a_2 + b_2i} =  \twoVec{(-1)^{f(00)}\left[a_1 + (-1)^{f(00) \oplus f(10)}b_1i\right]}{ a_2 + (-1)^{f(10) \oplus f(11)}b_2i}.
\end{equation}
In general, a particular algorithm, be it classical or quantum, solves a particular formulation of the problem;  i.e., it determines if a black-box of a \emph{specific type} which \emph{computes in some reasonable form} $f$ is balanced or constant.
Hence, it seems at least some of the apparent difference in powers of the classical and quantum solutions comes from the slightly different formulation of the problems, i.e., the different `powers' of the black-boxes.

A comment should be made about what it means to compute $f$ in some reasonable form. Since quantum computing requires unitarity, the simplest and perhaps most natural way to compute $f$ is with an $f$-controlled-NOT gate; indeed this was the original method proposed by Deutsch~\cite{Deutsch:1985aa}. 
One thus needs to be careful of claims that the separate output qubit for $U_f$ is not needed~\cite{Collins1998} and that the alternative quantum black-box $U_f'\ket{x}=(-1)^{f(x)}\ket{x}$ can equally well be used. Rather, it seems that $U_f'$ does not reasonably compute $f$ as absolute phase factors have no physical significance and it is hence impossible to characterise which boolean function $f$ the black-box `computes' by trying various inputs. 
The same issue is not present, however, in the classical de-quantised solution because phase factors are measurable in this case, and thus $C_f$ can be seen to compute $f$, albeit in a slightly non-standard way. 

\subsection{Black-box embeddings}




The quantum black-box (represented by $U_f$) is often considered an \textit{embedding} of the classical black-box computing $f$~\cite{Williams:2011uq}; if this were true it would be more reasonable to view the quantum solution as solving the original problem. 
This, however, is a misunderstanding which helps hide the differences between the classical and quantum formulations of the problem. 
For this to be an embedding we would require a map $e: \{0,1\}^n \hookrightarrow \mathcal{H}_{g(2^{n+1})}$ (where `$\hookrightarrow$' denotes an embedding) which preserves the structure of the computed function. That is, for $x \in \{0,1\}^n$, $e(f(x)) = U_f (e(x))$.
In other words, the computational states we assign to represent the binary bits 0 and 1 must behave as expected under $U_f$ given that $U_f$ should compute the function $f$.

However, the requirement of the unitarity of $U_f$ makes such an embedding impossible. 
This can be verified by considering any constant boolean $f$:  such a function is not bijective, so no bijective $U_f$ preserving the required structure can exist. 
Because no unitary embedding is possible,  it seems more suitable to consider the quantum solution as  a method to  solve an \textit{analogue} of the classical problem, rather than a more efficient solution to the classical problem. 
One is forced to conclude that the typical claims comparing the quantum and classical solution are, in fact, not valid. 
Any comparison of the problems should take into account the differences in complexity of the black-boxes~\cite{Abbott:2010ab}.

Interestingly, there is an embedding between the quantum black-box $U_f$ and the de-quantised black-box $C_f$, so it is not as unreasonable to compare the solutions using these black-boxes as it is to compare the quantum and one-dimensional classical solutions. 
By realising that the original classical solution and the de-quantised classical solution are not solving the same problem, we see there is no explicit contradiction with claims that $2^{n-1}+1$ black-box calls is the best that can be done in the original classical problem~\cite{Mermin:2007aa}.


\section{Algorithm Formulation}

Given a particular formulation of the problem, the algorithm formulation involves determining the input for the black-box, and what operations are required to determine the nature of $f$ from the output of the black-box.

In the standard quantum solution~\cite{Cleve:1997aa,Deutsch:1992aa}, we initially prepare our system in the state $\ket{00}\ket{1}$, and then operate on it with a three-qubit Hadamard gate, $H^{\otimes 3}$, to get:
\begin{equation}
\label{eqn:equalSuperpos}
H^{\otimes 3}\ket{00}\ket{1} = \frac{1}{2}\sum_{x\in \{0,1\}^2}\ket{x}\ket{-} = \ket{++}\ket{-}.
\end{equation}
After applying the $f$-controlled-NOT gate $U_f$ we have
\begin{equation}
\label{eqn:n2fcNot}
\begin{split}
	U_f \frac{1}{2}\sum_{x\in \{0,1\}^2}\ket{x}\ket{-} & = \sum_{x\in \{0,1\}^2}\frac{(-1)^{f(x)}}{2}\ket{x}\ket{-}.
\end{split}
\end{equation}
By applying a final 3-qubit Hadamard gate to project this state onto the computational basis we obtain the state
\begin{equation}
	(-1)^{f(00)} \ket{f(00) \oplus f(10)}  \otimes \ket{f(10) \oplus f(11)} \ket{1}.
\end{equation}
If we measure both the first and second qubits we can determine the nature of $f$: if both qubits are measured as $0$, then $f$ is constant, otherwise $f$ is balanced. This result is correct with probability one.

The de-quantised solution works in a similar way, but uses complex numbers as two-dimensional complex bits. Using the black-box $C_f$ defined previously, the algorithm proceeds as follows.
We set $z = z_1 =  z_2 = 1+i$, apply $C_f$ and multiply by $z$ to project onto the computational basis to obtain the result:
\begin{equation}
	\frac{z}{2}\times C_f \twoVec{z}{z} = \frac{1}{2}\times
	\begin{cases}
		\twoVec{(-1)^{f(00)}z^2}{z^2} = \twoVec{(-1)^{f(00)}i}{i} & \text{if $f$ is constant,}\\
		&\\
		\twoVec{(-1)^{f(00)}z\bar{z}}{z^2} = \twoVec{(-1)^{f(00)}}{i} &\\
		\twoVec{(-1)^{f(00)}z\bar{z}}{z\bar{z}} = \twoVec{(-1)^{f(00)}}{1} & \text{if $f$ is balanced.}\\
		\twoVec{(-1)^{f(00)}z^2}{z\bar{z}} = \twoVec{(-1)^{f(00)}i}{1} &\\
	\end{cases}
\end{equation}

By checking both of the resulting complex numbers, we can determine whether $f$ is balanced or constant with certainty. If both complex numbers are imaginary then $f$ is constant, otherwise it is balanced. In fact, the ability to determine if the output numbers are negative or positive allows us to determine the value of $f(00)$ and thus which Boolean function $f$ is; the quantum algorithm is incapable of doing this~\cite{Mermin:2007aa}.

The ability to de-quantise the $n=1$ and $n=2$ solutions suggests that, at least in these cases, the power of the quantum algorithm comes from exploiting the two-dimensionality of the black-box, rather than from quantum mechanical effects~\cite{Abbott:2011aa}. 

\section{Implementation of classical NMR computing}


\begin{figure}
\centering
\includegraphics[width=1\columnwidth]{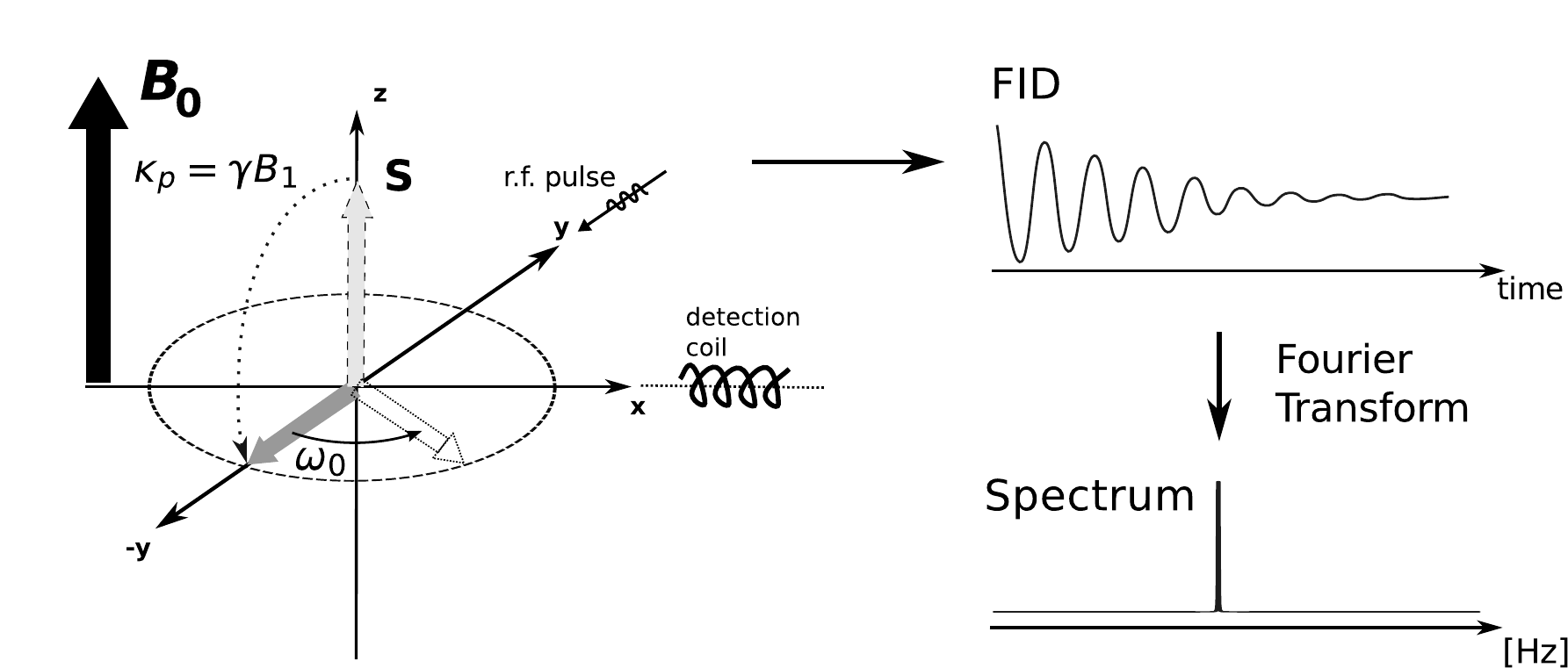}\caption{Creating and
observing a NMR signal. After reaching thermal equilibrium magnetisation in
the external magnetic field, the magnetisation vector $S$ is flipped to
the $xy$-plane of the laboratory frame of reference by applying a
radio-frequency (rf) pulse of suitable duration. After the rf pulse, the NMR
signal is detected in the $xy$-plane in the form of a time-domain
signal, the free induction decay (FID) which is recorded by a receiver coil,
here assumed to be placed in the $x$-direction. The FID is converted
into the frequency-domain spectrum by a Fourier transform.}%
\label{fig:magnetisation}%
\end{figure}

\begin{figure}
\centering
\includegraphics[width=.7\columnwidth]{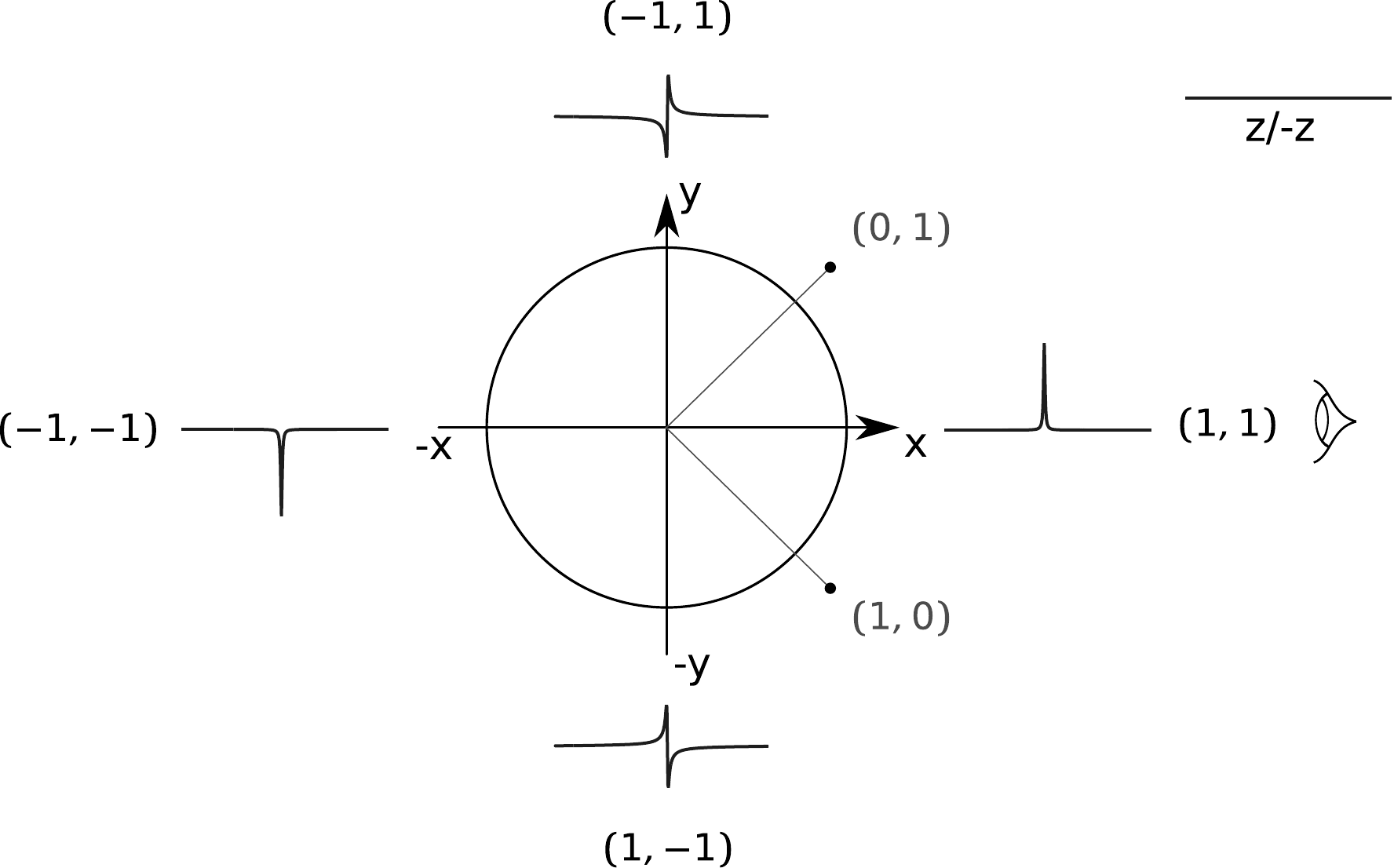}\caption{The embedding from classical
two-dimensional bits onto the nuclear magnetisation vectors in the $xy$-plane.
Assuming signal detection along the $x$-direction, the phases of the
corresponding NMR spectra are shown, together with the basis bits $(0,1)$ and
$(1,0)$ (see text).}
\label{fig:mapping}%

\end{figure}

For the implementation of the de-quantised algorithm we use real, two-dimensional vectors to represent our complex bits (only the direction of the vector is of particular importance); 
this is equivalent to using complex numbers, but more convenient for use in this implementation. 
In this representation our basis bits corresponding to the classical 0 and 1 become $(1,0)$ and $(0,1)$ respectively. 
For the implementation we must embed our bits into the physical medium; 
to do so, we use the nuclear magnetisation spin vectors for this embedding. 
Specifically, we embed the complex bits into the $xy$-plane of the rotating frame of reference in a NMR experiment
as is illustrated in Fig.~\ref{fig:magnetisation}. We only consider uncoupled spin species with
spin quantum number  $S=1/2$, and we have the advantage that every direction
of the magnetisation vector in the $xy$-plane is distinguishable. Working solely 
with uncoupled spin species, their dynamics are fully described by a classical model of 
magnetisation vectors subjected to a range of differnt rotations (pulses) \cite{Levitt:2008fk,Merino-maths}. Hence, at no point do we make explicit use of the quantum-mechanical properties of nuclear spin systems.
In particular, we take (row-vectors represent complex bits, column vectors are
the nuclear spin vectors; we omit normalisation factors for convenience):
\[
(1,0)\rightarrow I_{-45}=%
\begin{pmatrix}
1\\
\frac{1}{\sqrt{2}}(1-i)
\end{pmatrix}
,\qquad(0,1)\rightarrow I_{+45}=%
\begin{pmatrix}
1\\
\frac{1}{\sqrt{2}}(1+i)
\end{pmatrix}
.
\]
Combinations of the complex bits are naturally taken by the vector addition
(i.e. the embedding is linear) of the corresponding spin vectors (e.g.~
$(1,1)\rightarrow I_{x}=\tbinom{1}{1} $). This mapping is shown in Fig.~\ref{fig:mapping}. 
We emphasise that while the same nuclear spin state vectors are used as in conventional quantum computing experiments, we choose a different embedding of our algorithm into this spin-state vector space which is more flexible for use with complex bits.

\begin{figure}
\centering
\includegraphics[width=.5\columnwidth]{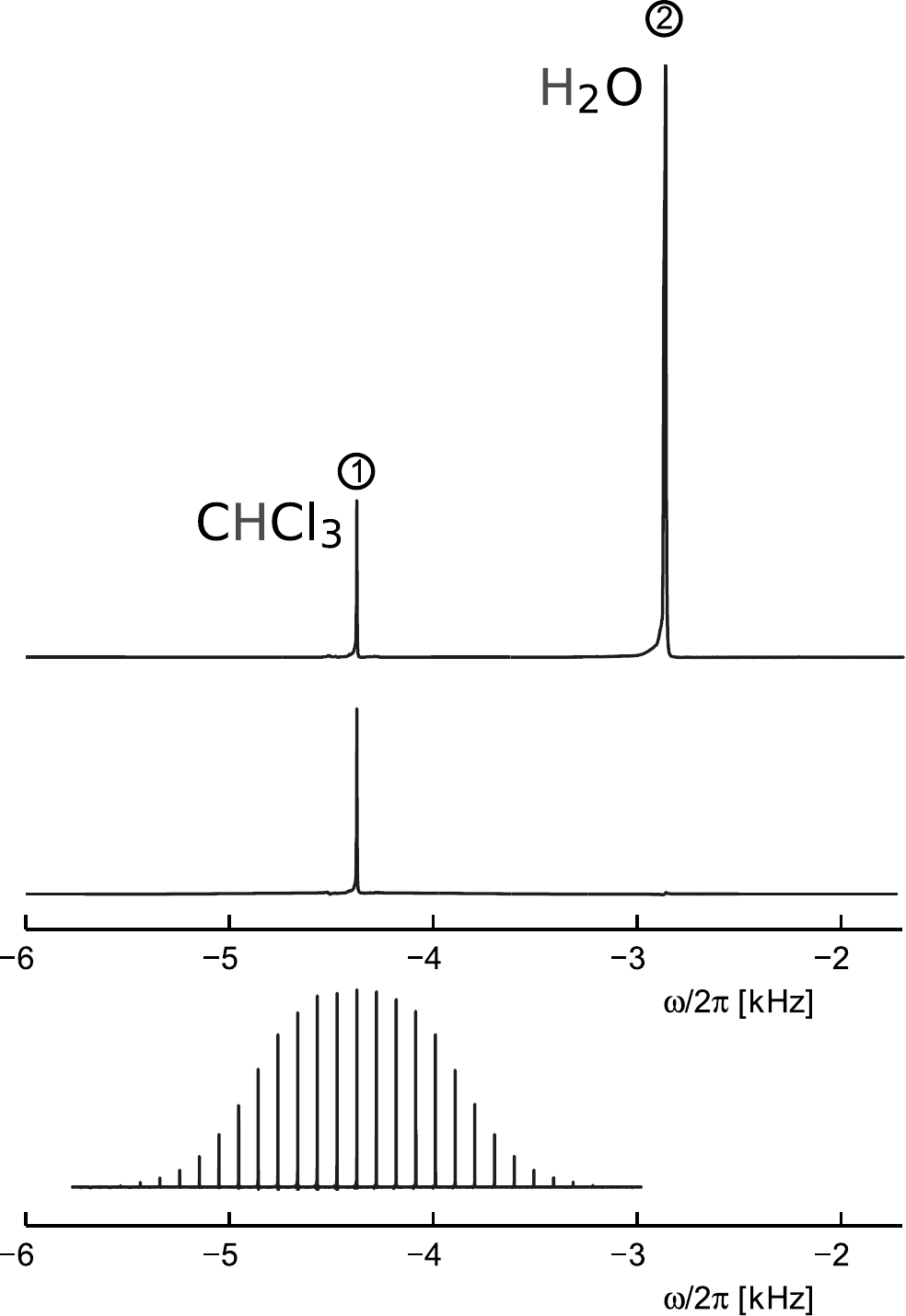}\caption{$^{1}$H NMR
spectrum ($\omega_{0}/2\pi=-600.13$ MHz) of 99.8\% deuterated chloroform with
a small amount of H$_{2}$O added (top trace). The middle trace shows the
$^{1}$H NMR spectrum after application of a selective 90\textdegree{} pulse,
centered around the resonance frequency of the CHCl$_{3}$ $^{1}$H NMR signal.
The bottom trace displays the excitation profile of the selective pulses used.
}%
\label{fig:spectra}%
\end{figure}

Our sample consists of 99.8\% deuterated chloroform with a small amount of
H$_{2}$O added. The $^{1}$H spins in the residual CHCl$_{3}$ and H$_{2}$O in
this mixture are used for the implementations. The top row in Fig.~\ref{fig:spectra} depicts
a conventional $^{1}$H NMR spectrum of the sample. The implementation for
$n=1$ only requires one spin species to be present. This is most easily
achieved by using selective excitation pulses, centred at the resonance
frequency of the CHCl$_{3}$ $^{1}$H NMR resonance. The $^{1}$H NMR spectrum
obtained by selective excitation, together with the corresponding excitation
profile of the selective pulses used are shown in the middle and bottom traces
of Fig.~\ref{fig:spectra}.

\subsection{The $n=1$ implementation}

\begin{figure}
\centering
\includegraphics[width=.6\columnwidth]{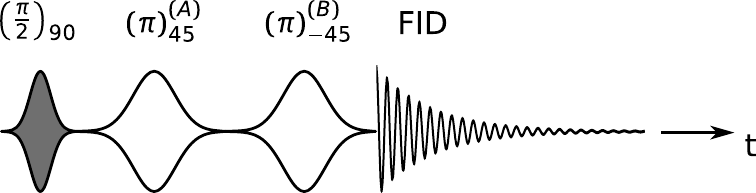}
\caption{The NMR pulse
sequence for implementation of the $n=1$ case;\ the flip angles effected by a
rf pulse are given in fractions of $\pi$, the symbols $A$, $B$ refer to the
control parameters of the black-box, curved shapes of rf pulses indicate
selective pulses~\cite{Freeman1998}.}%
\label{fig:n1seq}%
\end{figure}

The $n=1$ implementation relies on the form of the black-box $C_f$ operating as
$$C_f((a_1,b_1)) = ((-1)^{f(0)}a_1,(-1)^{f(1)}b_2).$$
The NMR pulse sequence for implementing this for our sample is
shown in Fig.~\ref{fig:n1seq}. The sequence starts with a (selective) $\pi/2$ pulse,
flipping the (CHCl$_{3}$) $^{1}$H magnetisation vector from the initial
equilibrium $z$-direction into the $xy$-plane, followed by a sequence of two
$\pi$ pulses applied to the $xy$-magnetisation vector as required to implement
the black-box before the resulting signal is detected in the form of a FID.

The four possible boolean functions for $n=1$ may be written $f_{AB}$ where $A=f(0)$
and $B=f(1)$. The four corresponding black-boxes can be uniformly implemented as the
following set of $\pi$ pulses, applied to our basis bits (see Fig.~\ref{fig:mapping}):
\[
(\pi)_{45}^{A}(\pi)_{-45}^{B}.
\]
We see that the physical embedding of the black-box fulfils the requirement
that the function $f$ is reasonably computed due to the ability to distinguish
all directions of magnetisation vectors in the $xy$-plane, and thus all valid
complex bits. In particular, applying the black-boxes for inputs $(1,0)$ or $(0,1)$
yields $((-1)^{f(0)},0)$ and $(0,(-1)^{f(1)})$ respectively as desired. The
effect of the four black-boxes on the `basis bits' is shown in
Table~\ref{tbl:n1BasisResult}, both in terms of complex bits and spin
vectors. 

\begin{table}
\centering
\begin{tabular}[c]{cc|cl}%
\toprule
$A=f(0)$ & $B=f(1)$ & $C_{f}((1,0))$ & $C_{f}((0,1))$\\
\midrule
0 & 0 & $(\phantom{-}1,0)$, $I_{-45}$ & $(0,\phantom{-}1)$, $I_{45}$\\
0 & 1 & $(\phantom{-}1,0)$, $I_{-45}$ & $(0,-1)$, $I_{-135}$\\
1 & 0 & $(-1,0)$, $I_{135}$ & $(0,\phantom{-}1)$, $I_{45}$\\
1 & 1 & $(-1,0)$, $I_{135}$ & $(0,-1)$, $I_{-135}$\\
\bottomrule
\end{tabular}
\caption{The effect of the four black-boxes on the basis-bit inputs for
$n=1$.}%
\label{tbl:n1BasisResult}
\end{table}

For the actual algorithm, the de-quantised solution can be simplified by
dropping the use of any equivalent of the final Hadamard operation; there is
no need to project the result onto the basis states since non-basis states are
equivalently detectable. In fact, the choice of $I_{\pm45}$ states as our
`basis bits' (rather than $I_{x}$ and $I_{y}$) is convenient because this
leaves the system in $I_{x}$ or $I_{y}$ states after the black-box, and these
states are particularly easy to distinguish (see Fig.~\ref{fig:mapping}).

\begin{figure}
\centering
\includegraphics[width=.8\columnwidth]{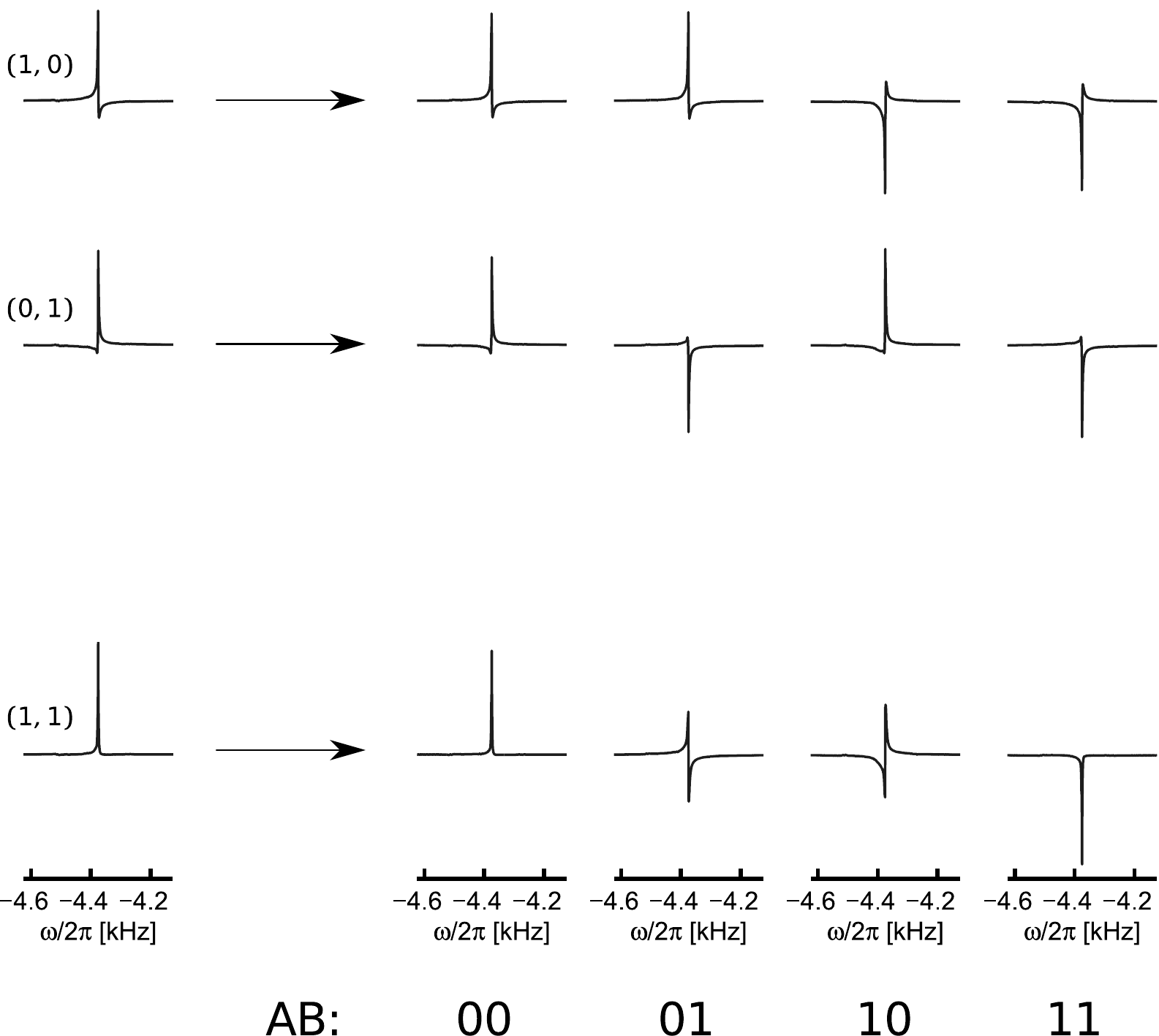}
\caption{$^{1}$H NMR spectra, CHCl$_{3}$ resonance,
implementing the $n=1$ case. The top two rows show the action of the black-box
on the basis-state input. The bottom row shows the computation itself, i.e.
with (1,1) input.}%
\label{fig:n1comp}%
\end{figure}

The effect of the algorithm with each black-box is shown in
Table~\ref{tbl:n1AlgResult}. The corresponding spectra relating to the
computation are shown in Fig.~\ref{fig:n1comp}, along with the action of the
black-box on the basis-state inputs. In particular for a constant $f$ we have
$I_{\pm x}$ spectra, and for balanced we have $I_{\pm y}$ spectra. Further,
because we can distinguish $I_{+x}$ spectra from $I_{-x}$ spectra (and
$I_{+y}$ spectra from $I_{-y}$ spectra) we can easily determine which function
$f$ the black-box represents. 

\begin{table}
\centering
\begin{tabular}[c]{cc|ccl}%
\toprule
$A=f(0)$ & $B=f(1)$ & Initial State & $(\pi/2)_{y}$ & $C_{f} = (\pi)_{45}^{A}(\pi)_{-45}%
^{B}$\\
\midrule
0 & 0 & $I_{z}$ & $(1,1)$, $I_{x}$ & $(\phantom{-}1,\phantom{-}1)$, $I_{x}$\\
0 & 1 & $I_{z}$ & $(1,1)$, $I_{x}$ & $(\phantom{-}1,-1)$, $I_{-y}$\\
1 & 0 & $I_{z}$ & $(1,1)$, $I_{x}$ & $(-1,\phantom{-}1)$, $I_{y}$\\
1 & 1 & $I_{z}$ & $(1,1)$, $I_{x}$ & $(-1,-1)$, $I_{-x}$\\
\bottomrule
\end{tabular}
\caption{The algorithm run with each of the four black-boxes for $n=1$.}%
\label{tbl:n1AlgResult}
\end{table}

\subsection{The $n=2$ implementation}

\begin{figure}
\centering
\includegraphics[width=.8\columnwidth]{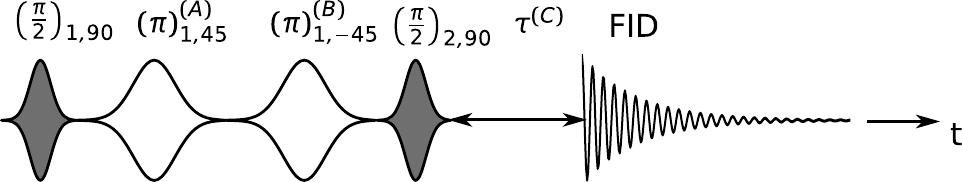}
\caption{The NMR pulse sequence
for implementing the $n=2$ case with two different spin species, 1 and
2;\ $\tau^{(C)}$ denotes a pre-acquisition delay of suitable duration.}%
\label{fig:n2seq}%
\end{figure}

The idea is that the $n=2$ case should be implemented as a natural extension
of the $n=1$ case by expanding it to include two different spin species. The
function to be implemented is:
\[
C_{f}((a_{1},b_{1})(a_{2},b_{2}))=\left(  a_{1}(-1)^{f(00)},b_{1}%
(-1)^{f(10)}\right)  \left(  a_{2},b_{2}(-1)^{f(10)\oplus f(11)}\right)  .
\]
With the same mapping of bits as for the $n=1$ case, the natural extension of
the embedding is to have the black-box, now defined by the three parameters $A=f(00)$, $B=f(10)$ and $C=f(10)\oplus f(11)$, implemented as:
\[
(\pi)_{1,45}^{A}(\pi)_{1,-45}^{B}(\pi)_{2,-45}^{C}.
\]
With this we can see that the first spin is treated as for the $n=1$ case, and
the second spin requires only a single pulse (or none, depending on $f$). This
form of the black-box, however, is only valid if one assumes ideal pulses. In
reality, we now deal with two different spin species with different resonance
frequencies, hence the receiver can only be on resonance with one species at a
time. In addition, all rf pulses have non-vanishing durations (in particular
the rather `soft' selective pulses with durations of the order of ms) during
which evolution of magnetisation occurs. Hence, the pulse sequence of the
physical implementation as shown in Fig.~\ref{fig:n2seq} has to account for non-ideal
behaviour in order to produce an equivalent result to the one obtained from
the idealised mathematical description. 

The NMR experiment starts with selective pulses centred on spin species 1
(exactly as for the $n=1$ case before), followed by a selective $\pi/2$ pulse
centred on spin species 2, and a delay of duration $\tau^{(C)}$. If the
receiver is kept on resonance with species 1 throughout, it is off resonance
for species 2. Accordingly, after the selective $\left(  \pi/2\right)  _{2}$
pulse the magnetisation of this species will precess in the $xy$-plane, and by
this acquire a phase difference relative to the on-resonance signal of species
1. This precession frequency depends on the difference in resonance frequency
between the two spin species and hence one can easily calculate the correct
duration of the delay $\tau^{(C)}$ that corresponds to a dephasing by
$3\pi/2$ from the starting $I_{x,2}$ condition to $I_{-y,2}$. At this point in
time, the situation is exactly equivalent to the ideal-pulse scenario and the
data acquisition is started. 

The pulse sequence depicted in Fig.~\ref{fig:n2seq} is the simplest form in which the
required idealised black-box can be carried out in a real NMR\ experiment. For
example, one might have started the pulse sequence with a non-selective
(`hard') $\pi/2$ pulse covering both spin species simultaneously. That would
be a common preparation step for a spin system, creating a common initial
state from which to work. However, this would only complicate matters as now
one would have to take into account dephasing of species 2 magnetisation
during all selective pulses applied to species 1, and one would have to apply
the corresponding phase corrections to any pulses applied to species 2 later
in the sequence. We do not \textit{have} to start from one common initial
state; instead our physical implementation avoids all unnecessary such
corrections and calculations by simply using selective pulses only and
preparing species 2 `just in time' such that only the duration of the
pre-acquisition delay needs to be calculated.

\begin{figure}
\centering
\includegraphics[width=1\columnwidth]{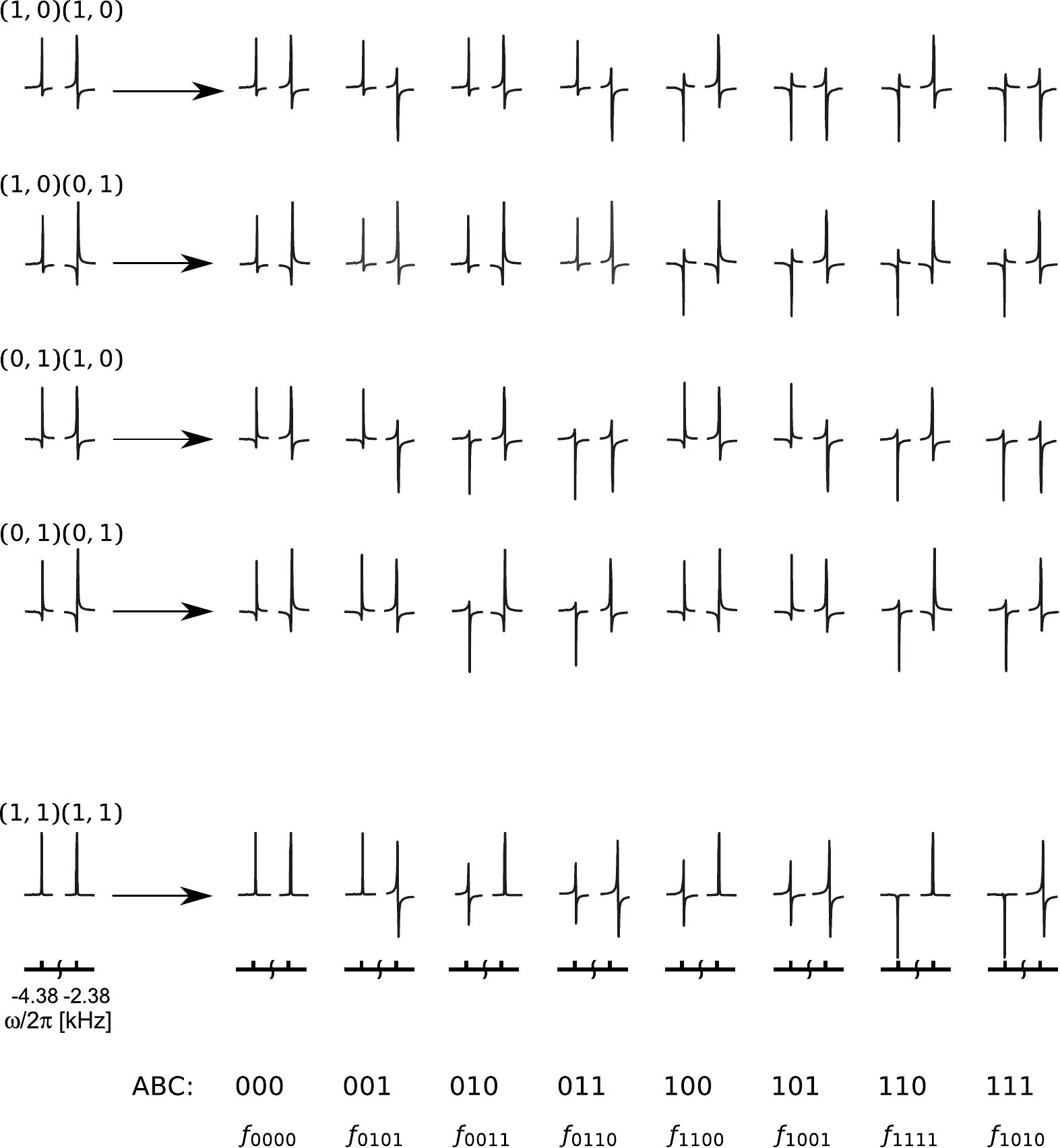}
\caption{$^{1}$H NMR spectra, CHCl$_{3}$ and H$_{2}$O
resonances, implementing the $n=2$ case. The top four rows show the action of
the black-box on the basis-state input. The bottom row shows the computation
itself, i.e. with (1,1) (1,1) input.}%
\label{fig:n2comp}%
\end{figure}

As for the $n=1$ case, it is easily verified that once again the requirements for the black-box to reasonably compute $f$ are satisfied: on a given `basis' input an even number of sign
changes indicates a function value of $0$ on this input, and odd number a
value of $1$. The results of the black-box on basis inputs is shown in
Table~\ref{tbl:n2BasisResult}, and the result of the algorithm is shown in
Table~\ref{tbl:n2AlgResult}. The corresponding input/output spectra are shown,
as for the $n=1$ case, in Fig.~\ref{fig:n2comp}. 

\begin{table}
	\centering
	\begin{tabular}{c|ccccc}
        \toprule
		A  B  C & $f$ & $C_f((1,0)(1,0))$ & $C_f((1,0)(0,1))$ & $C_f((0,1)(1,0))$ & $C_f((0,1)(0,1))$\\
		\midrule
		0  0  0 & $f_{0000}$ & $(\phantom{-}1,0)(1,0)$ & $(\phantom{-}1,0)(0,\phantom{-}1)$ & $(0,\phantom{-}1)(1,0)$ & $(0,\phantom{-}1)(0,\phantom{-}1)$\\
		0  0  1 & $f_{0101}$ & $(\phantom{-}1,0)(1,0)$ & $(\phantom{-}1,0)(0,-1)$           & $(0,\phantom{-}1)(1,0)$ & $(0,\phantom{-}1)(0,-1)$\\
		0  1  0 & $f_{0011}$ & $(\phantom{-}1,0)(1,0)$ & $(\phantom{-}1,0)(0,\phantom{-}1)$ & $(0,-1)(1,0)$           & $(0,-1)(0,\phantom{-}1)$\\
		0  1  1 & $f_{0110}$ & $(\phantom{-}1,0)(1,0)$ & $(\phantom{-}1,0)(0,-1)$           & $(0,-1)(1,0)$           & $(0,-1)(0,-1)$\\
		1  0  0 & $f_{1100}$ & $(-1,0)(1,0)$           & $(-1,0)(0,\phantom{-}1)$           & $(0,\phantom{-}1)(1,0)$ & $(0,\phantom{-}1)(0,\phantom{-}1)$\\
		1  0  1 & $f_{1001}$ & $(-1,0)(1,0)$           & $(-1,0)(0,-1)$                     & $(0,\phantom{-}1)(1,0)$ & $(0,\phantom{-}1)(0,-1)$\\
		1  1  0 & $f_{1111}$ & $(-1,0)(1,0)$           & $(-1,0)(0,\phantom{-}1)$           & $(0,-1)(1,0)$           & $(0,-1)(0,\phantom{-}1)$\\
		1  1  1 & $f_{1010}$ & $(-1,0)(1,0)$           & $(-1,0)(0,-1)$                     & $(0,-1)(1,0)$           & $(0,-1)(0,-1)$\\
        \bottomrule
	\end{tabular}
	\caption{The effect of the eight black-boxes on the basis-bit inputs for $n=2$.}
	\label{tbl:n2BasisResult}
\end{table}

\begin{table}
\centering
\begin{tabular}[c]{ccc|ccc}%
    \toprule
A & B & C & $f$ & $(\pi/2)_{y}$ & $C_{f} = (\pi)^{A}_{45,1}(\pi)^{B}%
_{-45,1}(\pi)^{C}_{-45,2}$\\
\midrule
0 & 0 & 0 & $f_{0000}$ & $(1,1)(1,1)$ & $(\phantom{-}1,\phantom{-}1)(1,\phantom{-}1)$\\
0 & 0 & 1 & $f_{0101}$ & $(1,1)(1,1)$ & $(\phantom{-}1,\phantom{-}1)(1,-1)$\\
0 & 1 & 0 & $f_{0011}$ & $(1,1)(1,1)$ & $(\phantom{-}1,-1)(1,\phantom{-}1)$\\
0 & 1 & 1 & $f_{0110}$ & $(1,1)(1,1)$ & $(\phantom{-}1,-1)(1,-1)$\\
1 & 0 & 0 & $f_{1100}$ & $(1,1)(1,1)$ & $(-1,\phantom{-}1)(1,\phantom{-}1)$\\
1 & 0 & 1 & $f_{1001}$ & $(1,1)(1,1)$ & $(-1,\phantom{-}1)(1,-1)$\\
1 & 1 & 0 & $f_{1111}$ & $(1,1)(1,1)$ & $(-1,-1)(1,\phantom{-}1)$\\
1 & 1 & 1 & $f_{1010}$ & $(1,1)(1,1)$ & $(-1,-1)(1,-1)$\\
\bottomrule
\end{tabular}
\caption{The algorithm run with each of the eight
black-boxes for $n=2$. While the idealised version of the black-box pulses for
species 2 is quoted here, the physically implemented pulse sequence gives
identical final results.}%
\label{tbl:n2AlgResult}
\end{table} 

\if01
\section{Comparison with quantum implementations}

There are a range of quantum implementations with varying properties, so it is not straight forward to decide how best to compare classical and quantum algorithms.

The main reason for the advantages of the classical implementation comes from the extra freedom we take in embedding the complex bits in the quantum system. In the quantum case, there is only freedom of embedding up to the choice of an orthonormal basis; any choice leads to the exact same disadvantage regardless, and is only for ease rather than anything else. Since our vectors representing complex bits contain only real components, we can embed them in the $xy$-plane of magnetisation. In this embedding global phases of the complex bit do not correspond to global phases of the quantum spin, so are detectable. In the classical computation we are still exploiting quantum spins via NMR, but doing so in such a way that the quantities we are interested in measuring behave classically. 

In practice, the quantum NMR algorithm should operate on a pseudo-pure mixed state, i.e. $\rho = (1-\varepsilon)2^{-n}I + \varepsilon \ket{\psi}\bra{\psi}$, which in general (for more than 1 qubit) takes some effort to prepare. However, for the Deutsch-Jozsa algorithm many implementations do not use such pseudo-pure states, and use the same initial state we use for the classical algorithm, i.e. a thermal state with an applied $B_z$ field. This means they are using a slightly different algorithm than the standard Deutsch-Jozsa algorithm; this is fine as long as the black-box embedding is the same that would be applied even if they had used a pseudo-pure state.

\if01
It seems desirable to be able to compare classical and quantum implementations, but what should we compare? Some possible things to consider are:
\begin{itemize}
 	\item What is the cost in preparing the initial state, how does this differ between the two?
 	\item What is the cost of having to deal with coupling? (this might not be so clear with respect to this algorithm, but our experiences can be used to talk about cost difference in general between quantum and classical NMR algorithms).
 	\item Is the quantum black-box more difficult to create than the classical one? We can use this to possibly compare the extra power in the black-box in different embeddings?
 	\item Is there a difference in difficulty in switching between functions computed between quantum/classical implementations? Is one more uniform?
 	\item Are there significant differences in times required?
 	\item Are there differences in accuracy or limitations between the methods?
 	\item Do the answers to any of these questions change between the $n=1$ and $n=2$ cases? Is there more difficulty in moving from $n=1$ to $n=2$ or above between classical and quantum cases?
\end{itemize}
\fi
\fi

\if01
\section{Classical and quantum NMR Computation}

- Emphasise the point that this is ensemble quantum computation which is not the same as normal QC, and has some important differences\\
- Review the fact the NMR is also suitable for classical comp\\

\fi

\if01
\section{Observations on quantum nature of NMR computation}

\begin{itemize}
	\item In general the pseudo-pure states used are not entangled for small enough $\varepsilon$, so what classifies it as quantum computation, especially in comparison to our classical computation where we are using quantum systems to perform classical computation?
 	\item In order to do controlled operations even with the classical implementation, coupling would be needed. This then starts to sound and behave very much like the quantum implementation. So, a) what makes our implementation classical rather than quantum (and vice-versa)? b) is there an argument for doing pseudo-quantum computation using our embedding? The underlying quantum basis we use is not orthonormal, but the basis is orthonormal w.r.t. the complex bits. What is the fundamental importance of an orthonormal quantum basis? Only the orthonormal states are the eigenvalues of the observables, i.e. they are the measurable ones. But in expectation value quantum computation (e.g. NMR based QC), this is not the case and we can detect anywhere in the $xy$-plane. Why? This needs to be clarified I think.
\end{itemize}
\fi

\section{Summary}

The experiments described in this paper not only show the successful implementation of the de-quantised Deutsch-Jozsa algorithm for $n=1,2$, but also bring attention to some important points regarding quantum computations. The experiments confirm that the Deutsch-Jozsa problem can indeed be solved classically in some cases using as many black-box calls as the standard quantum algorithm. 
During the process of developing our implementation some issues about the comparisons of algorithms for different formulations of the problem were raised and discussed in detail.
We note that, while we explicitly demonstrate the ability to solve the $n=1,2$ cases for what is essentially an alternative (but equivalent to the quantum) problem formulation, the ability to do so for higher values of $n$ reduces to the ability to de-quantise the quantum solution for these cases, which is believed not to be possible~\cite{Abbott:2011aa}.

These experiments reiterate the utility of NMR as a classical computing substrate~\cite{Rosello-Merino:2010aa}. 
Specifically, the use of an alternative embedding from the problem space onto the uncoupled nuclear spin vectors allowed us to perform the classical algorithm and even determine the specific function $f$ computed. This is a more general result regarding the ensemble computational nature of NMR computing, which allows any spin-direction in the $xy$-plane to be (in theory) resolvable. It is further possible that this alternative embedding could be utilised for quantum computations involving only real-valued coefficients. Since such computations are universal~\cite{Bernstein:1997fk}, it is plausible that this could be used to perform ensemble quantum computations in which the state amplitudes themselves can be measured directly.

\section{Experimental}

99.8\% deuterated $\mathrm{CHCl}_{3}$ was obtained from Aldrich Chemicals. $^{1}\mathrm{H}$ NMR spectra of a $\mathrm{CHCl}_{3}$~/~$\mathrm{H}_{2}\mathrm{O}$ mixture  were recorded on Bruker Avance II 600 NMR spectrometer (see Fig.~\ref{fig:spectra}), corresponding to a $^{1}\mathrm{H}$ Larmor frequency of $-600.13$~MHz. On-resonant 90\textdegree{} pulse durations were $2.5$~ms and Gaussian selective  pulses at a resolution of 1000 points utilised cut-offs of 1\%. All spectra were recorded in single scans, allowing recycle delays of at least 11~s between experiments.

\section{Acknowledgments}
We gratefully acknowledge the support of this work by the Deutsche Forschungsgemeinschaft (MB and AS) and EPSRC (AA, MB and AS). We thank Susan Stepney for comments which helped improve the paper.

\bibliography{NMRde-quant}

\begin{thebibliography}{10}

\bibitem{Abbott:2011aa}
A.~A. Abbott.
\newblock (2011).
\newblock The {D}eutsch-{J}ozsa problem: De-quantisation and entanglement.
\newblock {\em Natural Computing}, to appear.

\bibitem{Abbott:2010ab}
A.~A. Abbott and C.~S. Calude.
\newblock (2010).
\newblock Understanding the quantum computational speed-up via de-quantisation.
\newblock {\em EPTCS}, 26:1--12.

\bibitem{Arvind2001a}
Arvind.
\newblock (2001).
\newblock Quantum entanglement and quantum computational algorithms.
\newblock {\em Pramana---Journal of Physics}, 56:357--365.

\bibitem{Merino-maths}
M.~Bechmann, A.~Sebald, and S.~Stepney.
\newblock Boolean logic gate design principles in unconventional computers: an
  {NMR} case study.
\newblock accepted: Int. J. Unconventional Computing.

\bibitem{Bernstein:1997fk}
E.~Bernstein and U.~Vazirani.
\newblock (1997).
\newblock Quantum complexity theory.
\newblock {\em SIAM Journal on Computing}, 26(5):1411--1473.

\bibitem{Calude:2007aa}
C.~S. Calude.
\newblock (Jun 2007).
\newblock De-quantizing the solution of {D}eutsch's problem.
\newblock {\em International Journal of Quantum Information}, 5(3):409--415.

\bibitem{Cleve:1997aa}
R.~Cleve, A.~Ekert, C.~Macchiavello, and M.~Mosca.
\newblock (Jan 1997).
\newblock Quantum algorithms revisited.
\newblock {\em Proceedings of the Royal Society of London Series A},
  1998(454):339--354.

\bibitem{Collins1998}
D.~Collins, K.~W. Kim, and W.~C. Holton.
\newblock (Sep 1998).
\newblock {D}eutsch-{J}ozsa algorithm as a test of quantum computation.
\newblock {\em Physical Review A}, 58(3):R1633--R1636.

\bibitem{Collins2000}
D.~Collins, K.~W. Kim, W.~C. Holton, H.~Sierzputowska-Gracz, and E.~O.
  Stejskal.
\newblock (Jul 2000).
\newblock {NMR} quantum computation with indirectly coupled gates.
\newblock {\em Physical Review A}, 62(2):022304.

\bibitem{Cory2000}
D.~G. Cory, R.~Laflamme, E.~Knill, L.~Viola, T.~F. Havel, N.~Boulant,
  G.~Boutis, E.~Fortunato, S.~Lloyd, R.~Martinez, C.~Negrevergne, M.~Pravia,
  Y.~Sharf, G.~Teklemariam, Y.~S. Weinstein, and W.~H. Zurek.
\newblock (2000).
\newblock {NMR} based quantum information processing: Achievements and
  prospects.
\newblock {\em Fortschritte der Physik}, 48(9-11):875--907.

\bibitem{Das2003}
R.~Das and A.~Kumar.
\newblock (SEP 2003).
\newblock Use of quadrupolar nuclei for quantum-information processing by
  nuclear magnetic resonance: Implementation of a quantum algorithm.
\newblock {\em Physical Review A}, 68(3):032304.

\bibitem{Deutsch:1985aa}
D.~Deutsch.
\newblock (Jan 1985).
\newblock Quantum theory, the {C}hurch-{T}uring principle and the universal
  quantum computer.
\newblock {\em Proceedings of the Royal Society of London Series A},
  400:97--117.

\bibitem{Deutsch:1992aa}
D.~Deutsch and R.~Jozsa.
\newblock (Jan 1992).
\newblock Rapid solution of problems by quantum computation.
\newblock {\em Proceedings of the Royal Society of London Series A},
  439:553--558.

\bibitem{Dorai2000}
K.~Dorai, Arvind, and A.~Kumar.
\newblock (Mar 2000).
\newblock Implementing quantum-logic operations, pseudopure states, and the
  {D}eutsch-{J}ozsa algorithm using noncommuting selective pulses in {NMR}.
\newblock {\em Physical Review A}, 61(4):042306.

\bibitem{Fahmy2008}
A.~F. Fahmy, R.~Marx, W.~Bermel, and S.~J. Glaser.
\newblock (AUG 2008).
\newblock Thermal equilibrium as an initial state for quantum computation by
  {NMR}.
\newblock {\em Physical Review A}, 78(2, Part A):022317.

\bibitem{Fitzsimons2007}
J.~Fitzsimons, L.~Xiao, S.~C. Benjamin, and J.~A. Jones.
\newblock (JUL 20 2007).
\newblock Quantum information processing with delocalized qubits under global
  control.
\newblock {\em Physical Review Letters}, 99(3):030501.

\bibitem{Freeman1998}
R.~Freeman.
\newblock (1998).
\newblock Shaped radiofrequency pulses in high resolution {NMR}.
\newblock {\em Progress in Nuclear Magnetic Resonance Spectroscopy},
  32:59--106.

\bibitem{Gopinath2008}
T.~Gopinath and A.~Kumar.
\newblock (2008).
\newblock Implementation of controlled phase shift gates and {Collins} version
  of {D}eutsch-{J}ozsa algorithm on a quadrupolar spin-7/2 nucleus using
  non-adiabatic geometric phases.
\newblock {\em Journal of Magnetic Resonance}, 193(2):168--176.

\bibitem{Kessel2002}
A.~R. Kessel and N.~M. Yakovleva.
\newblock (DEC 2002).
\newblock Implementation schemes in {NMR} of quantum processors and the
  {D}eutsch-{J}ozsa algorithm by using virtual spin representation.
\newblock {\em Physical Review A}, 66(6):062322.

\bibitem{Kim2000}
J.~Kim, J.-S. Lee, S.~Lee, and C.~Cheong.
\newblock (Jul 2000).
\newblock Implementation of the refined {D}eutsch-{J}ozsa algorithm on a
  three-bit {NMR} quantum computer.
\newblock {\em Physical Review A}, 62(2):022312.

\bibitem{Kumar2002}
A.~Kumar, K.~V. Ramanathan, T.~S. Mahesh, N.~Sinha, and K.~V.~R. Murali.
\newblock (AUG 2002).
\newblock Developments in quantum information processing by nuclear magnetic
  resonance: Use of quadrupolar and dipolar couplings.
\newblock {\em Pramana---Journal of Physics}, 59(2, Sp. Iss. SI):243--254.
\newblock 2nd Winter Institute on Foundations of Quantum Theory and Quantum
  Optics, Kolkata, India, JAN 02-11, 2002.

\bibitem{Levitt:2008fk}
M.~H. Levitt.
\newblock (2008).
\newblock {\em Spin Dynamics: Basics of Nuclear Magnetic Resonance}.
\newblock John Wiley \& Sons, 2nd edition.

\bibitem{Linden1998}
N.~Linden, H.~Barjat, and R.~Freeman.
\newblock (1998).
\newblock An implementation of the {D}eutsch-{J}ozsa algorithm on a three-qubit
  {NMR} quantum computer.
\newblock {\em Chemical Physics Letters}, 296(1-2):61--67.

\bibitem{Mahesh2001}
T.~S. Mahesh, K.~Dorai, Arvind, and A.~Kumar.
\newblock (2001).
\newblock Implementing logic gates and the {D}eutsch-{J}ozsa quantum algorithm
  by two-dimensional {NMR} using spin- and transition-selective pulses.
\newblock {\em Journal of Magnetic Resonance}, 148(1):95--103.

\bibitem{Mangold2004}
O.~Mangold, A.~Heidebrecht, and M.~Mehring.
\newblock (OCT 2004).
\newblock {NMR} tomography of the three-qubit {D}eutsch-{J}ozsa algorithm.
\newblock {\em Physical Review A}, 70(4):042307.

\bibitem{Marx2000}
R.~Marx, A.~F. Fahmy, J.~M. Myers, W.~Bermel, and S.~J. Glaser.
\newblock (Jun 2000).
\newblock Approaching five-bit {NMR} quantum computing.
\newblock {\em Physical Review A}, 62(1):012310.

\bibitem{Mermin:2007aa}
N.~D. Mermin.
\newblock (2007).
\newblock {\em Quantum Computer Science}.
\newblock Cambridge University Press.

\bibitem{Rosello-Merino:2010aa}
M.~Rosell{\'o}-Merino, M.~Bechmann, A.~Sebald, and S.~Stepney.
\newblock (2010).
\newblock Classical computing in nuclear magnetic resonance.
\newblock {\em International Journal of Unconventional Computing},
  6(3--4):163--195.

\bibitem{Wei2003}
D.~X. Wei, J.~Luo, X.~P. Sun, X.~Z. Zeng, X.~D. Yang, M.~L. Liu, and S.~W.
  Ding.
\newblock (FEB 2003).
\newblock {NMR} experimental realization of seven-qubit {D-J} algorithm and
  controlled phase-shift gates with improved precision.
\newblock {\em Chinese Science Bulletin}, 48(3):239--243.

\bibitem{Williams:2011uq}
C.~P. Williams.
\newblock (2011).
\newblock {\em Explorations in Quantum Computing}.
\newblock Springer-Verlag, New York, Inc., second edition.

\end{thebibliography}

\end{document}